# How Informal Science Education Influences Elementary Students' Perceptions of Science and Themselves

Molly K. Finn[1], Renato Mazzei[1], Blake Drechsler[1], Zoie Telkamp[1], Mihika Rao[1], Prakamya Agrawal[2], and Anne McAlister[3]

[1]Department of Astronomy, University of Virginia, Charlottesville, VA; [2]SLAC National Lab, Stanford University, Palo Alto, CA; and [3]Department of Engineering Education, University at Buffalo, Buffalo, NY

Keywords: Elementary Students, Astronomy, Student Identity, Nature of Science, Out-of-school Time

**ABSTRACT:** Underrepresentation in STEM fields starts early, with elementary students already showing differences based on gender and race in their interest in science, belief in their ability to do science, and belief that their personal identity aligns with being a scientist. Here we discuss an out-of-school time astronomy program that focuses on enriching science education in under-served school systems to promote students' excitement about science and help them see scientists as people just like them. Before, after, and throughout the program, we survey students on their perceptions of science, themselves, and their activities. We find that over the course of our program, students become more confident in their science abilities. Student ideas about science remain unchanged, but largely align with Nature of Science ideals. We also find that on days that students report they were creative and asked questions, they were more likely to say they felt like a scientist and were interested in the day's topic. Our results suggest that incorporating creativity and opportunities to ask questions can be just as important as doing experiments for generating interest in and a sense of participating in science.

# INTRODUCTION

Despite decades of effort to improve diversity in science careers, women and people of color remain largely underrepresented, especially in the physical sciences (Fry et al., 2021). In a study of children in the 10-18 age range in the UK, the majority of those surveyed admire scientists and find science interesting, but only 16% aspired to become scientists (Archer et al., 2020). This difference between students' science interest and science aspiration, which the authors refer to as the "being/doing" divide (Archer et al., 2010), is more pronounced for female students than for male students. In this study we focus on the factors that influence elementary students' science aspirations, including their interest in science, their self-efficacy, and how their identity aligns with being a scientist.

Many studies indicate that science interest is high and shows no difference between genders for younger students, but a gap in interest between male and female students develops around grade six (Pell and Jarvis, 2001; Murphy and Beggs, 2005; Baram-Tsabari and Yarden, 2011). Several studies also show that around this age, female students begin to have a weaker belief in their abilities to do science and math (Andre et al., 1999; Gunderson et al., 2012; Bian et al., 2017). The differences in elementary student science interest and self-efficacy by ethnicity/race are not as prominent as by gender (Archer et al., 2020; Vandenberg et al., 2021), although significant achievement gaps exist (Quinn and Cooc, 2015; Curran and Kellogg, 2016) as well as aspiration gaps (DeWitt and Archer, 2015). There is much less work done examining the role of socioeconomic status in elementary student science aspirations, but Archer et al. (2020) do find that students from socio-economically disadvantaged families are significantly less likely to aspire to science careers.

Studies find that differences in science aspirations around this age may be due in part to students becoming more aware of cultural stereotypes that make them believe their personal identity is incompatible with being a scientist (Schreiner and Sjøborg, 2007; Ceci et al., 2009; Archer et al., 2010; Cheryan et al., 2015; Carli et al., 2016). This is noted both based on gendered stereotypes (Vincent-Ruz and Schunn, 2018) as well as ethnic/racial stereotypes (Achbacher et al., 2010; Archer et al., 2015; Rahm et al., 2022). Students may enjoy science and consider themselves good at science, but still think that science is "not for me." Many aspects of the scientist stereotype are inherently contradictory to society's concepts of being feminine (Archer et al., 2010) or to students' perceptions of their cultural identity (Archer et al., 2015). To encourage more students from underrepresented populations to consider pursuing science careers, science educators need to address all of these different effects in the way that science is presented to children, especially in the age range where they begin to develop their personal identities.

Traditionally, many prominent depictions of science and scientists in the media (Stienke, 2005), which are often shared and perpetuated by parents and elementary teachers (Keller, 2001), reinforce ideas about science being a series of facts to be learned rather than a process, or that science is only done by "brilliant" white men in lab coats. Representation in science has improved over the last several decades, both in the media and in real life (Long et al., 2010; Steinke et al., 2017; Previs, 2016), but children are still more likely to draw a man than a woman when asked to draw a scientist (Miller et al., 2018; Hayes et al., 2020). When asked to draw themselves doing science, fifth grade students often drew themselves reading a book or taking notes, while the scientists they drew were mostly white men in a laboratory setting (Barman et al., 1997). While student perceptions of themselves doing science may have improved since the Barman et al. (1997) study, incorporating hands-on, interactive activities into science curricula has been shown to help students develop positive attitudes towards science and better understand the many different ways that science happens (Bredderman, 1983; Aydede and Kesercioğlu, 2010; Satterthwait, 2010).

Another means of improving students' perceptions of science and enhancing their interest in science is to teach them about nature of science (NOS) (Tobias, 1990; van Griethuijsen et al., 2014). Here we use the definition from Schwartz et al. (2004, pg. 611) that NOS is "the values and underlying assumptions that are intrinsic to scientific knowledge, including the influences and limitations that result from science as a human endeavor", and which are characterized by the following features from the National Science Teaching Association (NSTA, 2000):

1. Scientific knowledge is reliable, yet still tentative.
2. Science uses a variety of methods.
3. Science involves creativity.
4. Science investigates questions related to the natural world.
5. The terms "theories" and "laws" have specific meanings in science.
6. Contributions to science have been made by people all over the world.
7. Science occurs in a social and cultural context.
8. The history of science shows science can both gradually and suddenly change.
9. There is a relationship between science and technology, but basic scientific research is not concerned with practical outcomes.

In this study, we examine the influence of the astronomy out-of-school time (OST) program Dark Skies, Bright Kids (DSBK), which targets elementary students from underrepresented populations. OST programs have been shown to impact student perceptions of science, scientists, and their own science identity, and they provide an excellent opportunity for them to learn science more interactively (Krishnamurthi and Porro, 2008; Bhattacharyya et al., 2011; McCreedy and Dierking, 2013; Riedinger and Taylor, 2016; Hayes et al., 2020). We investigate the impact of DSBK programs on the students' perceptions of science, NOS concepts, and their own self-identity in relation to science.

## PROGRAM BACKGROUND

**Organization Overview.** Dark Skies, Bright Kids (DSBK) is a graduate-student-run volunteer organization based at the University of Virginia (UVA). DSBK was founded in 2008 to provide OST outreach opportunities for elementary-aged children in underserved communities - primarily in Charlottesville, Virginia and the greater surrounding area, but also across the state of Virginia. The majority of DSBK's members are graduate student volunteers in the UVA Astronomy Department, which includes the authors of this paper. Though our outreach activities are astronomy-themed, our mission is more generally aimed toward fostering the curiosity and imagination of students through hands-on, group-oriented science activities. We aim to provide children with the opportunity to see science as fun and exciting, and to see themselves as scientists, through engagement in inquiry-based learning activities that occur outside the traditional classroom environment.

**Program Structure.** The two main ways that DSBK interacts with our target demographic are through after-school clubs and week-long summer clubs. For the after-school variant, each semester (i.e., twice per year, once in the spring and once in the fall) we coordinate with a single elementary school in the City of Charlottesville or Albemarle County to host a club. Schools are selected on a rotating basis, with priority given to schools with a large population of students from backgrounds underrepresented in STEM fields (especially non-white or low socioeconomic status). Typical after-school clubs meet for approximately two hours on Friday afternoons for eight weeks. Each week a particular topic in astronomy serves as the theme for the day, and we lead the students through several activities related to the topic. To provide the students with a varied learning experience, we do different kinds of activities such as demonstrations, hands-on interactive activities, and physically active kinesthetic activities. The week-long summer clubs are led and structured much the same way, but with the content delivered in a single week. Summer camps are conducted in rural areas of Virginia and are available to students from any local elementary school. We cover the same daily topics and activities in the form of ten half-day sessions (morning and afternoon) from Monday to Friday.

Typically, all DSBK events are done in-person, with the exception of Summer 2020, Spring 2021, and Summer 2021, which were operated virtually due to COVID. To facilitate compatibility with the remote learning environment, we modified several aspects of camp, such as rewriting some activities so they could be safely performed at home with minimal supervision, providing students with pre-packaged materials and instructions, and introducing the content and activity descriptions in synchronous video sessions. The video sessions gave students the opportunity to interact with DSBK volunteers in real time and ask questions as they worked through the activities. We made recordings of the video sessions available so that students had an asynchronous option in cases of poor internet connection or unreliable device availability. A more thorough discussion of the structures and differences between virtual and in-person DSBK events can be found in Finn et al. (2020).

## METHODS

**Participants.** We began collecting the data presented here in 2019 from students who participated in a DSBK program between Fall 2019 and Summer 2021. In that time period, there were a total of three semester after-school clubs and five summer clubs offered, for a total of 135 students with at least some data. The Spring 2020 semester club was cut short due to COVID-19, so no after-club surveys were collected. All club participants were elementary students between 3rd and 5th grade, or in the case of the summer clubs, rising 3rd-

Table 1. *Summary of data sources from the seven clubs conducted during the data collection period.*

| Club Date | Format | Number of students with data | Number of before-club surveys | Number of daily surveys | Number of after-club surveys |
|---|---|---|---|---|---|
| 2019 Summer | In-person | 10 | 10 | 77 | 10 |
| 2019 Summer | In-person | 12 | 12 | 79 | 12 |
| 2019 Fall | In-person | 21 | 21 | 99 | 18 |
| 2020 Spring | In-person | 16 | 15 | 16 | 0 |
| 2020 Summer | Virtual | 34 | 34 | 124 | 18 |
| 2021 Spring | Virtual | 10 | 10 | 47 | 8 |
| 2021 Summer | Virtual | 9 | 9 | 48 | 6 |
| 2022 Summer | In-person | 8 | 8 | 50 | 8 |

6th graders, corresponding roughly to ages 8-12. In Table 1, we summarize the data obtained from each club.

Elementary students were asked to apply to participate in the programs. For semester clubs, we gave applications to all students within the eligible grade window at the school, and for summer clubs we used online advertisements and provided copies of the applications to local elementary schools. The number of applicants rarely exceeded the capacity of the club, but in those cases applicants were randomly selected (the number of participants for each club, which is almost always the number of applicants, is shown in Table 1). This means that our sample is primarily composed of students who chose to join our astronomy OST program. Most often this is because the student was already interested in science or astronomy, but there could be other reasons that a student or their guardian signed them up, such as wanting a free after school program or summer camp to provide adult supervision. We discuss the ways this selection affects our results in the Limitations section.

There was a total of 135 students with at least some data across all the programs analyzed in this work. We included on the applications the option to disclose demographic information for the students. Of those who included this data, 47% were female, 35% were non-white, and 29% qualified for free or reduced lunch (a measure of socioeconomic status). The number of students who identified as Black, Hispanic, Asian, or a combination of two or more races is too small for individual group statistics, and so in this analysis we have grouped them together.

**Data Sources.** Data was collected in the form of surveys that students filled out at different stages during the club. We developed these surveys to test student beliefs about NOS concepts, their science ability and aspiration, if they identified with being a scientist during out programs, and how different daily activities correlated with their science identity. We designed all of the questions to be simple and easily understood by children, and all forms were limited to one page to accommodate short attention spans. Each of these surveys included a variety of qualitative (short answer) and quantitative questions (Likert scale, true false, or check mark). In this work, we consider only the responses given to the quantitative questions. [Full versions of the survey forms are available from the corresponding author, upon request.] There were three types of surveys:

- **Before-club survey ("Getting to Know You")**: This form was filled out by participants once, at the beginning of the first day of club. The quantitative questions diagnosed the students' perceptions of their science ability, their creativity, their aspiration to become a scientist, and their beliefs about NOS concepts.

- **Daily survey ("Wrap-Up")**: This form was filled out by participants at the end of each full day of the club. For an eight-week semester club, each student was given eight wrap-up surveys over the course of the club, while for a week-long summer camp, each student was given five wrap-up surveys. Each of these asked quantitative questions about if they felt like a scientist that day, were interested in that day's topic, if they had any questions about the topic, and what activities they performed throughout the day (such as asking questions or doing experiments).

- **After-club survey ("Saying Goodbye")**: This form was filled out by participants once, at the end of the last day of club. The quantitative questions asked were identical to those on the before-club survey.

Table 2. *Summary of the quantitative questions asked on the before- and after-club survey, given at the beginning of the first day of club and the end of the last day of club, respectively. For the True or False statements, the bolded and underlined answer is the one that better aligns with NOS concepts.*

| Question | Answer Format |
|---|---|
| Do you feel like you are good at science? | Likert scale, 1-5 |
| Do you feel like you are creative? | Likert scale, 1-5 |
| When you grow up, do you want to be a scientist? | Likert scale, 1-5 |
| Science is based on observations. | **True** or False |
| There is a right way to do science. | True or **False** |
| Science requires creativity. | **True** or False |
| Teamwork is part of science. | **True** or False |
| Science can change. | **True** or False |
| Science teachers do science. | **True** or False |
| Science is challenging. | True or **False** |

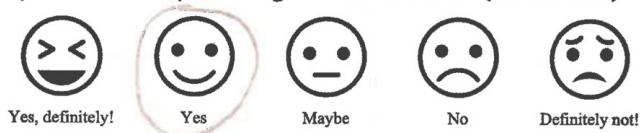

Figure 1. An example of one of the questions on the before-club survey as it was presented to the students in graphical format. For our analysis, student answers to these Likert scale questions were translated to a 1 to 5 scale, where "Yes, definitely!" = 5 and "Definitely not!" = 1.

To help promote student engagement with the Likert scale questions, we presented the possible answer choices as a kid-friendly graphic, shown in Figure 1. In Tables 2 and 3, we list the full set of questions analyzed in this paper for the before/after-club survey and the daily survey, respectively.

We include in the before- and after-club surveys state-ments about NOS concepts written in kid-friendly language. There is no one definition of NOS (Abd-El-Khalick and Lederman, 2000), but in Table 2 we highlight in bold the answer that better aligns with NOS principles listed in the introduction (NSTA, 2000) and which we aimed to convey throughout the program. During DSBK clubs, NOS concepts are not explicitly mentioned or taught, but were implicitly modeled in our activity design and discussions. Entering this work, we hypothesized that participation in a DSBK club would improve students' understanding of NOS, and that this would be reflected in the after-club survey.

## DATA ANALYSIS AND RESULTS

In this section, we address three main research questions:

1. How did the students' perceptions of their science ability, creativity, and science aspirations change before and after the astronomy club?

2. How did the students' perceptions of the NOS concepts change before and after the astronomy club?

3. Which daily activities were correlated with students feeling like a scientist or interested in the day's topic?

**RQ1: How did the students' perceptions of their science ability, creativity, and science aspirations change before and after the astronomy club?** To investigate this question, we compared responses on the before- and after-club surveys to the questions "Do you feel like you are good at science?", "Do you feel like you are creative?", and "When you grow up, do you want to be a scientist?". We matched before and after responses and only included in this analysis students who had completed both surveys, for a total of 74 pairings. We show the mean response to the three questions before and after the astronomy club in Figure 2. The responses to each question are mostly positive, especially the first two questions measuring their perceived science competency and creativity, for which there were no responses of "No" or "Definitely not". Their science aspirations were on average lower than the other two questions, which aligns with previous findings (e.g. Archer et al., 2020), but still were on

**Table 3**. *Summary of the quantitative questions asked on the daily surveys, given at the end of each day of club.*

| Question | Answer Format |
|---|---|
| Today I was creative. | Check mark |
| Today I asked questions. | Check mark |
| Today I had fun. | Check mark |
| Today I made observations. | Check mark |
| Today I did experiments. | Check mark |
| Today I worked with others. | Check mark |
| Today I learned something. | Check mark |
| Today I changed my mind about something. | Check mark |
| Today I felt like a scientist. | Likert scale, 1-5 |
| [Today's activity] is interesting. | Likert scale, 1-5 |

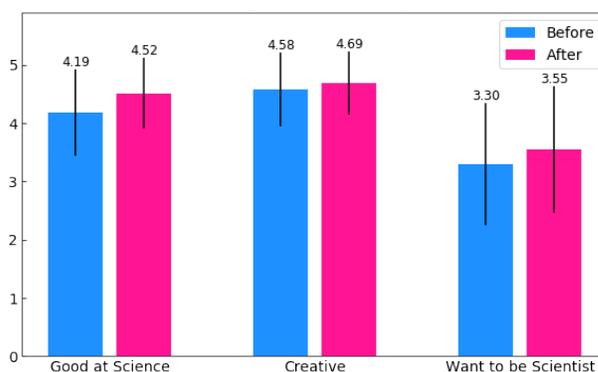

**Figure 2.** Mean and standard deviation of before- and after-club responses to the questions "Do you feel like you are good at science?", "Do you feel like you are creative?", and "When you grow up, do you want to be a scientist?", where 5 is the most positive response.

average more positive than negative or neutral.

A Shapiro-Wilk test for normality revealed that before and after data for each question did not follow a normal distribution (p<0.01 for all 6 tests). We then used a non-parametric Wilcoxon signed-ranked test to show that the before and after responses did not change significantly over the course of our programs, at the 95% confidence threshold for the questions "Do you feel like you are creative?" (p = 0.096) and "When you grow up, do you want to be a scientist?" (p = 0.082). For the question "Do you feel like you are good at science?", students' perceptions of their science ability significantly increased after participation in our program at the 95% confidence threshold (p = 0.0034). This indicates that our program positively impacted students' perceptions of their science abilities.

*Demographic Effects.* We investigated how these differences varied for students based on their gender, race, and socioeconomic status. We compared before- and after-club responses to each of the questions with Wilcoxon signed-rank tests for subsets of participants who self-reported as male, female, white, non-white, qualifying for free-or-reduced lunch (FRL), and not qualifying for FRL. Since some participants chose not to disclose one or more of these demographics, these subsets do not reflect the total number of students with before- and after-club survey pairings.

The average response to the question "Do you feel like you are good at science?" increased after the program for each demographic subset, but was only statistically significant based on a 95% confidence threshold for female students, white students, and students who do not qualify for FRL. We show the mean response to the question "Do you feel like you are good at science?" for each demographic subset in Figure 3. The Wilcoxon signed-rank test indicated that there was no significant difference between responses for the questions "Do you feel like you are creative?", and "When you grow up, do you want to be a scientist?", therefore we do not show those here.

We also performed Mann–Whitney U tests, a non-parametric test to determine if the average responses to each question were consistent between gender, race, and FRL status. Using a 95% confidence threshold for significance, we find that male students have a significantly higher average response to "Do you feel like you are good at science?" than female students before club, but not after club. There is also a significant difference between white and non-white students' responses to that question before club, but not after club, with non-white students having more positive responses than white students. We find no significant difference between students who qualify for FRL and those who do not and no differences in either of the other questions. These test results are summarized in Table 4.

The lack of statistical significance in some of these cases could be due to the smaller sample sizes involved when we isolate demographic groups (the minimum size is N=18 for students who qualify for FRL). While the Mann-Whitney U test and Wilcoxon signed rank test are still valid for small sample sizes, they are only sensitive to larger differences. In the case of comparing the Mann-Whitney U tests before and after club, the sample sizes for each group have not changed. This means that while there was a statistical difference in male vs. female and white vs. non-white student responses before club, after club those differences at least decreased enough that they were no longer statistically significant.

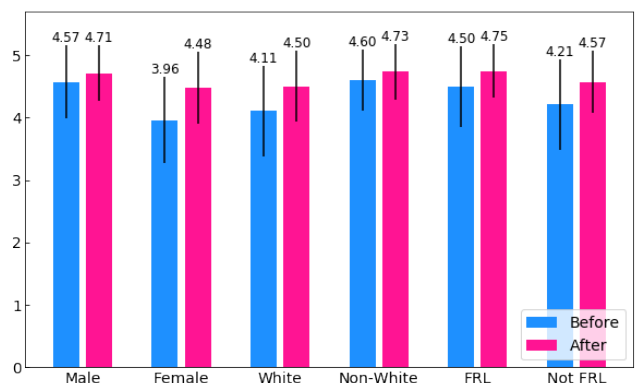

**Figure 3.** Mean and standard deviation of before- and after-club responses to the question "Do you feel like you are good at science?", where 5 is the most positive response, broken up into demographic subsets.

**Table 4.** *Results of the Mann–Whitney U test to determine if there is a significant difference in responses to "Do you feel like you are good at science?" between demographic groups before and after the club. Pairings with a significant difference based on a 95% confidence threshold (p < 0.05) are highlighted in green. There were no pairings with a significant difference for either of the other questions on the survey.*

| Comparison Groups | Mann–Whitney U test *p*-value |
| --- | --- |
| Male/Female before club | 0.0022 |
| Male/Female after club | 0.088 |
| White/Non-white before club | 0.017 |
| White/Non-white after club | 0.10 |
| FRL/Not before club | 0.13 |
| FRL/Not after club | 0.15 |

**Table 5.** *Fraction of responses that were "True" for each NOS statement. These are averaged between before and after clubs with a 95% confidence interval reported.*

| Nature of Science Statement | Fraction of 'True' Responses and Error |
| --- | --- |
| Science is based on observations. | 0.91±0.11 |
| There is a right way to do science. | 0.14±0.11 |
| Science requires creativity. | 0.89±0.13 |
| Teamwork is part of science. | 0.95±0.11 |
| Science can change. | 0.88±0.13 |
| Science teachers do science. | 0.87±0.12 |
| Science is challenging. | 0.70±0.16 |

**RQ2: How did the students' perceptions of NOS concepts change before and after the astronomy club?** We next compared before- and after-club responses to the series of True/False questions concerning NOS concepts. Figure 4 shows the fraction of responses that were "True" before and after the club. All of the statements show a small change before and after the club. Most of those changes are in the expected direction (see Table 2), with the exception of "There is a right way to do science", "Teamwork is part of science", and "Science can change." To determine whether the changes in responses were statistically significant, we computed a 95% confidence interval for each. Considering this confidence interval, none of the observed changes in the fraction of "True" responses are statistically significant. We also looked at these responses as a function of gender, race, and socioeconomic status and found no significant differences or changes before and after club.

The lack of a statistically significant change in NOS perceptions aligns with other findings that an implicit teaching approach is not effective in teaching NOS concepts (Abd-El-Khalick and Lederman, 2000). While we cannot comment further on the effect of the astronomy club on changing students' perceptions, it is still interesting to consider their overall perception of science. We show the fractions of "True" responses averaged between before and after club with respective errors in Table 5. The majority of students demonstrated perceptions that align with NOS concepts for every statement, notably including the ideas that science requires creativity, that there is more than one right way to do science, and that science can change, which are some of the more common misconceptions about NOS (Lederman et al., 2002). Students mostly recognized the collaborative nature of science, with the highest fraction of agreement with the statement that "Teamwork is a part of science."

We expected that students would come to view science as less challenging through participation in our programs. However, this statement elicited the largest spread in student responses. We know that after the program, 98% of students responded with either "Yes" or "Yes, definitely!" in response to the statement "I am good at science", so it seems that the belief that science is challenging does not come from them feeling that they struggle with science or science class, but rather a recognition of the inherent challenge in science.

**RQ3: Which daily activities were correlated with students feeling like a scientist or interested in the day's topic?** To answer this question, we looked at student responses to the daily "Wrap-Up" surveys. Students checked boxes of whether or not they performed various activities that day during camp, and responded to the Likert scale questions "Today, I felt like a scientist." and "[The day's topic] is interesting." We look for a relationship between students performing a given activity and a higher score in either of those two Likert scale questions.

To quantify such a relationship, we used a proportional odds logistic regression model using the `polr()` function from the `MASS` package in R. This model analyzes how different factors affect the odds that a student will respond with a higher value on a Likert scale question. We input into this model the responses from each day's wrap-up form from each student, where we have a total of 540 completed forms from a total of 124 students. We looked for a dependence of the ordinal responses to the two Likert scale questions on whether students reported performing an activity, represented in the model as a 1 or a 0. We also included in the model a dependence on which program the response is from to control for variations in student personalities, the format of the program, and the volunteers leading the program.

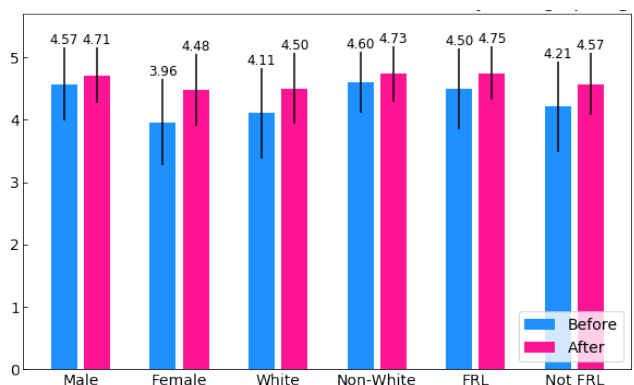

**Figure 4.** Fraction of student responses that were "True" for each NOS statement, before and after the astronomy club. While all of the statements show a small change in responses, the error bars indicate that none of the changes are significant based on a 95% confidence threshold.

To determine that the above inputs to the model resulted in the best fit of the data, we minimized the Akaike Information Criterion (AIC) score, where a lower score indicates a better model fit. For example, including which program the response came from resulted in a lower AIC than not including the club information. Other parameters we considered including, but which resulted in higher AIC values, were whether the program was a week-long summer camp or a semester-long weekly after school club, whether the program was virtual or in-person, and which topic was being covered that day. The first two of these options are both accounted for by including which program the response is from as an input parameter. The fact that the model fits better without including the day's topic indicates that it is not correlated with how the students responded to the two Likert scale questions.

The outputs from the model are the odds ratios and p-values for each input. Together, these indicate which of the various inputs resulted in a statistically significant higher-level ordinal response. So in our case, the model outputs the odds that when a student reported doing a given activity, they would also give a more positive response to "Today, I felt like a scientist" or "[The day's topic] is interesting". An odds ratio above 1 indicates a positive association, while an odds ratio below 1 indicates a negative association. The results for each of the questions are shown in Table 6.

Both questions had four factors with a statistically significant relationship with the responses. For the responses to "Today, I felt like a scientist", those four factors were being creative, doing experiments, asking questions, and making observations. All of these factors had a positive relationship with how much students felt like a scientist. Being creative had the strongest positive relationship with the responses.

For the responses to "[The day's topic] is interesting", being creative, asking questions, and having fun had positive associations with the students' responses, while changing their mind had a negative relationship. Again, being creative was the strongest of these.

## DISCUSSION

Overall, our data indicate that students felt more confident about their science abilities after participating in a DSBK program. This finding aligns well with other studies evaluating the effects of OST programs (Chun and Harris, 2011; McCreedy and Dierking, 2003). Of the 74 paired before and after responses, only three students responded to "Do you feel like you are good at science?" with "Maybe" at the end of the program, and all other responses were "Yes" or "Definitely yes". While 70% said that "Science is challenging", 94% of the students said that they were good at science, which indicates that while students acknowledge the difficulty of science, they do not think that it is too difficult for them to excel at the subject.

We do not see similar increases in students' responses to the questions "Do you feel like you are creative?". This is unsurprising since it is not a skill that we emphasize in our program or a skill that is traditionally associated with science. However, all but four students responded with either "Yes" or "Definitely yes" to "Do you feel like you are creative?" before the program, so the lack of difference before and after club may also be because there was little room for improvement in the first place.

We also do not see a statistically significant increase in responses to the question "Do you want to be a scientist when you grow up?". This is also unsurprising and agrees with prior work that even when students have positive attitudes towards science, a much smaller fraction aspire to be scientists when they grow up (Archer et al, 2020).

When considering the influence of student demographics, we find that female students, white students, and students who do not qualify for free-or-reduced lunch showed the greatest increase in science confidence over the course of DSBK programs. This could potentially be due in part to the fact that a majority of our volunteer base is white and female, and so students with this identity are exposed to representation among the scientists that they meet through the program. Future work could include tracking volunteer demographics more closely to investigate this connection. We were surprised to see that before club, non-white students had higher science confidence responses than their white counterparts, based on the history of whiteness in STEM disciplines. This could be due in part to the selection effect of which students decide to join DSBK programs. The fact that we combined several racial groups together also makes this result harder to interpret, though it aligns with recent findings from Archer et al., (2020) that among 10-19 years old in the UK, South-Asian, Chinese, and Black students had higher science aspirations than white students.

**Table 6.** *Results of the proportional odds logistic regression model to determine the odds that performing an activity resulted in the student reporting a higher ordinal response to the Likert scale question. We highlight activities that have a statistically significant correlation with the Likert scale responses based on a 95% confidence threshold. The final column shows how often students reported doing each activity.*

| "Today, I…" | "Today, I felt like a scientist" | | "[The day's topic] is interesting" | | % of responses |
|---|---|---|---|---|---|
| | *p*-value | Odds ratio | *p*-value | Odds ratio | |
| Was creative | <0.01 | 3.11 | <0.01 | 2.19 | 74% |
| Did experiments | <0.01 | 2.25 | 0.90 | 0.97 | 72% |
| Asked questions | <0.01 | 1.83 | 0.014 | 1.85 | 56% |
| Worked with others | 0.78 | 0.94 | 0.58 | 0.86 | 60% |
| Had fun | 0.97 | 0.99 | 0.045 | 1.90 | 87% |
| Learned something | 1.00 | 1.00 | 0.32 | 1.34 | 79% |
| Made observations | 0.040 | 1.66 | 0.29 | 1.33 | 67% |
| Changed my mind about something | 0.079 | 0.68 | <0.01 | 0.51 | 42% |

We find that in the daily surveys, students who said that on a given day they were creative or that they asked questions were more likely to also say that they felt like a scientist and that they were interested in that day's topic. Being creative had a particularly strong association, with this activity having the greatest positive relationship with responses to both questions. We found this result surprising, since creativity is not traditionally associated with being a scientist and is more often associated with artistry and perceived as the opposite of science (Kind and Kind, 2007). We know that scientists do require creativity though, which is included in our NOS understanding (NSTA, 2000), and the students in this study also recognized this idea, as demonstrated by 89% of them agreeing with the statement "Science requires creativity."

A commonality between creativity and asking questions is that both are a form of having ownership in what you are learning. Being able to impart some of their personal ideas into the activities can make students feel more like a part of the process, and so feel more like they are scientists. Similarly, asking questions requires internalizing what you have learned and either identifying what you still do not understand or thinking of what new information would be interesting. Both of these activities can make the students feel more engaged with the material and like they have agency in what they are learning. This may be why students were more interested in the topic and more likely to feel like they were scientists when they partook in creative activities or asked questions.

It is less surprising, but still important to note, that doing experiments and making observations were both also correlated with students feeling like scientists. Many of the activities that we do during DSBK programs are led with a focus on experimentation, especially when leading activities that would otherwise be based on more passive demonstration. We present the activities to students with language that makes it explicitly clear that we are doing an experiment. One example of an activity that has worked well for our group is when we frame the Coke and Mentos demonstration as an experiment of whether diet or regular Coke creates a bigger reaction. This language can help the students recognize that the activity is an experiment and that they need to make observations to uncover the answer, which in turn can help them realize they are taking part in a scientific process. These experiments were correlated with students feeling more like scientists, but not with being more interested in the science topic.

It is similarly unsurprising that when students say that they had fun during the program, they were more likely to find the day's topic interesting. What is more surprising is that a student responding that they had fun during the day did not have a much stronger correlation with their interest in the day's topic than being creative or asking questions. This suggests then that students' interest in the learning material taught during the day is not their primary source of fun during the program.

We find no correlation with either feeling like a scientist or being interested in the topic for the statements "Today I learned something new" and "Today I worked with others". This is surprising to us since we would have expected that learning science topics would increase students' science identity and students generally agreed with the statement "Teamwork is a part of science" (95% respond "True"). This may be because students frequently work together in many contexts, and so while they recognize teamwork is important to science, teamwork is not inherently scientific on its own.

The fact that learning something new shows so little correlation with feeling like a scientist or being interested in the days' topics could be because learning new things can be positive, negative, or neutral and may also feel more like a traditional school activity. Students are expected to learn new things all the time in school whether they are interested in the topic or not and learning about science, math, or writing in school does not necessarily make them think of themselves at scientists, mathematicians, or writers. This result demonstrates how active learning and hands-on experiences can be more important for developing science identity than gaining knowledge about science.

While students for the most part recognize that science is tentative and subject to change (88% responded that the statement "Science can change" is True), they still seemed uncomfortable with changing their minds. On days that students reported changing their minds, the odds of them saying they were interested in the day's topic was half as much as on days they did not change their minds. Changing their minds also resulted in an odds ratio of less than one for responding positively to "Today I felt like a scientist", although that result is less statistically significant. This suggests that while students acknowledge that science chang-es, they may have been turned off from the topic when they themselves changed their minds. It is a natural reaction to be uncomfortable when confronted with something that changes your mind, and in our society changing one's mind often comes with strong negative connotations about being wrong. It would be helpful to find a framework for leading students through that process while maintaining their interest in and excitement for the topic. There have been many studies on what science misconceptions exist among elementary students (e.g. Stein et al. 2011; Soeharto et al., 2019) and how educators can address and correct those misconceptions (e.g. Gooding and Mets 2011; Karpudewan et al., 2017). However, there is much less literature on how correcting misconceptions affects students' interest in science or their confidence in their science abilities.

**Actionable Suggestions**. Below we list actionable suggestions for elementary OST science education:

- Incorporate student creativity into science activities.
- Encourage students to think of questions about what they have learned.
- Frame lessons or science demonstrations as experiments with observations and inferences.
- Aspects of activities that make a student feel more like a scientist do not necessarily make students more interested in the topic; carefully consider goals for an activity as you craft it.
- In cases where the goal of a program or lesson is to increase interest in science or to help students feel like scientists, focus on students having science experiences rather than learning science knowledge.
- If you are interested in teaching the nature of science, use explicit instruction on these topics.

## LIMITATIONS

The biggest limitation of this study is that the sample of students that participated all chose to be a part of the DSBK programs, and so were more likely to already have a favorable view of science and their science abilities. Furthermore, we only received daily responses from students who attended club on a given day, and if a student chose to stop coming to our program or missed the last day of club, they did not complete an after-club survey and so were not included in those comparative parts of the analysis (of the 118 students who completed before-club surveys, only 74 completed after-club surveys). This likely biased the responses to questions about students' perceptions of their science competency, science aspirations, daily science identity, and interest in the daily science topics. It may also have influenced their prior knowledge and perceptions of NOS concepts. This could be improved by instead conducting this study in a setting where all students in a class are expected to participate in the activities.

Similarly, students who experienced behavior problems during a day of the program and students who may be on the ADHA-spectrum are more likely to not complete our daily wrap-up survey forms. We do not collect data on these issues, and so they may bias our results on the daily surveys towards students with longer attention spans or those who had more positive experiences during the day.

A drawback of the check mark format that we used for the daily activities was that it allowed for straightlining and there were several cases of students checking every box. It is possible that the students thoughtfully believed they had engaged in all of the activities listed, but there were also likely many cases of the students checking every box without giving much thought to the exercise. Since we cannot meaningfully distinguish between those two cases, all data was included. To account for this in future iterations of this work, we could consider including an attention check to distinguish whether students are giving genuine responses to the surveys.

We find that our approach of implicitly demonstrating NOS concepts without explicit discussion of them throughout the program did not result in any significant change in student perceptions of those NOS ideas. To better understand how such programs could influence student understanding of NOS concepts, it would be helpful to incorporate explicit discussion of these concepts into our program.

## CONCLUSIONS AND SUMMARY

In this work, we present survey results from elementary students participating in DSBK astronomy programs, both in 8-week-long after school clubs and week-long full-day summer camps. Students were asked about their perceptions of their science competency, creativity, science aspirations, and understanding of NOS concepts both before and after the programs, and throughout the programs were asked about the activities they took part in, their perceptions of themselves as scientists, and their interest in the daily astronomy topic. Our major findings are summarized below:

- Students' confidence in their science abilities increased after participating in DSBK programs. Their belief in their creativity and their aspiration to pursue a scientific career also both increased, but not statistically significantly based on a 95% confidence threshold. Female students, white students, and students who do not qualify for free-or-reduced lunch showed the greatest increase in science confidence over the course of the programs.
- Before the program, male students and non-white

- students showed a significantly higher science confidence than female students and white students, respectively. After the club, there was no statistically significant difference between their responses. This was due to the increase in science confidence of female and white students.

- There was no statistically significant change in students' understanding of NOS concepts after the program, suggesting that an implicit teaching approach is insufficient to convey NOS ideas. However, the majority of students' beliefs did still align with accepted NOS tenets, both before and after the program.

- Being creative was the daily activity that had the strongest association with both students feeling like scientists and with students being interested in the day's topic. Students who said they had been creative on a given day were 3 times more likely to say they felt like a scientist and 2 times more likely to say they were interested in the day's topic.

- In addition to being creative, there was a positive association between feeling like a scientist and the daily activities of asking questions, doing experiments, and making observations. For interest in the day's topic, there was also a positive association with asking questions and having fun.

- There was no significant correlation between learning something new and either feeling like a scientist or being interested in the day's topic. The same is true for working with others.

- There was a negative correlation between students reporting that they had changed their mind about something and saying they were interested in the day's topic, with students half as likely to be interested on days that they changed their minds.

These conclusions can be actively incorporated into future STEM education programs by emphasizing student creativity and questioning in science activities, performing experiments and making observations, and taking an explicit approach to teaching NOS concepts.

## AUTHOR INFORMATION

**Corresponding Author**
Molly Finn, University of Virginia, 530 McCormick Rd, Charlottesville, VA 22904. Email: mf4yu@virginia.edu

**Author Contributions**
The manuscript was written through contributions of all authors. All authors have given approval to the final version of the manuscript. We have no known conflict of interest to disclose.

## ACKNOWLEDGMENTS
We thank Sean Finn for his insightful discussions and helpful advice on statistical analyses.

## FUNDING SOURCES
This material is based on work supported by the National Science Foundation Graduate Research Fellowship Program under grant No. 1842490. Any opinions, findings, and conclusions or recommendations expressed in this material are those of the authors and do not necessarily reflect the views of the National Science Foundation.

## ABBREVIATIONS
AIC: Akaike information criterion; DSBK: Dark Skies, Bright Kids; FRL: free or reduced lunch; NOS: nature of science; OST: out-of-school time; STEM: science, technology, engineering, mathematics; UVA: University of Virginia